\documentclass[]{revtex4}

\usepackage{graphicx}
\begin{document}

\title{Functional renormalization group in the two-dimensional Hubbard model}
\author{Carsten Honerkamp}
\affiliation{Max-Planck Institute for Solid State Research, \\
D-70569 Stuttgart, Germany \\ C.Honerkamp@fkf.mpg.de}
\date{November 10, 2004}

\begin{abstract}
We review recent developments in functional renormalization
group (RG) methods for interacting fermions. These
approaches aim at obtaining an unbiased picture of competing Fermi 
liquid instabilities in the low-dimensional models like the
two-dimensional Hubbard model. We discuss how
these instabilities can be approached from various sides and how
the fermionic RG flow can be continued into phases with broken
symmetry.
\end{abstract}

\maketitle

\section{Introduction}
Competing Fermi liquid instabilities are a general feature
of  low-dimensional interacting lattice electron systems.
These instabilities are driven by long-lived particle-particle or
particle-hole excitations near the Fermi surface (FS). 
They can lead to long range correlations of various types 
and a destruction of Fermi liquid behavior. 
In two and higher dimensions perturbative renormalization
group methods are probably the least biased approach to study
such systems at weak coupling and low energy scales in a transparent way.

In the last decade various functional renormalization group (fRG) methods
\cite{shankar,msbook,zanchi,halboth,salmhofer}
for many-fermion systems have been developed. 
These schemes are capable of capturing the interplay of various
low-energy scattering processes within certain approximations.
However, the practical experience  shows that
each scheme has advantages and disadvantages.
This has led to the development of alternative flow schemes\cite{tflow,ape}. 
They provide the possibility to tackle the
 competing instabilities from different angles.
Fortunately, it turns out that besides well-understood
conceptual differences the flow of these schemes agrees
quite favorably over a wider parameter range of models like the
two-dimensional Hubbard model. With this understanding we hope that 
further improvements of the approach are on a firm basis.

This short review is organized as follows. First we outline the
general formalism of the renormalization group for the
one-particle irreducible vertex functions\cite{wetterich,salmhofer,kopietz}. 
We explain how cutoff flow, temperature flow  ($T$-flow) and 
interaction flow (IF) are obtained. 
Then we review some results for the two-dimensional Hubbard model. 
Finally we describe
a novel approach how the fermionic RG flow can be continued into
phases with broken symmetry.

\section{Functional RG for 1PI vertex functions} \label{formalism}

\subsection{General setup}
The class of models we study here is described by an action of the form
\begin{equation}\begin{array}{ccl}
S (\psi,\bar\psi)& =&  \sum_{k,s} \,  \bar \psi_{k,s}
Q(k) \, \psi_{k,s} \\[2mm] &&
+ \frac{1}{2N}\, T^3 \, \sum_{k,k',q \atop
s,s'} V(k,k',k+ q)
\bar \psi_{k+q,s} \bar \psi_{k'-q,s'}
\psi_{k',s'}  \psi_{k,s} \, . \end{array}
\label{ham}  \end{equation}
Here $\bar \psi (k,s)$ and  $\psi (k,s)$ are Grassmann fields
representing fermions with wavevector $\vec{k}$, Matsubara
frequency $k_0$ (we write $k=(k_0, \vec{k})$) and spin projection
$s=\pm 1/2$. $Q(k)$ denotes the quadratic part of the action,
here given by
$
Q(k)= T \left[ -i k_0 + \epsilon (\vec{k})  \right] 
$. 
For fermions on a 2D square lattice with nearest neighbor hopping $t$, next-nearest neighbor hopping $t'$ and chemical potential $\mu$, the dispersion is
\begin{equation} 
\epsilon (\vec{k}) = -2t \left(  \cos k_x + \cos k_y \right) - 4 t'\cos k_x \cos k_y - \mu \, .\end{equation} 
$V(k,k',q) $ defines the spin-rotationally invariant, frequency-
and wavevector-conserving interaction between two fermions. For the
repulsive onsite interaction of the Hubbard model $V(k,k',q) =U>0$.

The connected correlation functions of the theory defined by
(\ref{ham}) can be obtained by taking derivatives of the
generating functional,
\begin{equation}
e^{-W(\xi,\bar \xi)} = \int {\cal D}\psi(k,s) {\cal D}\bar \psi
(k,s) \, e^{-S(\psi,\bar\psi) + \sum_{k,s} \left[ \bar \psi(k,s)
\xi (k,s) + \bar \xi(k,s) \psi (k,s) \right]}
\label{expW} \, . \end{equation}
For our purposes it is simpler to work with vertex functions,
which are generated by the Legendre transform of $W(\xi , \bar \xi)$ with respect to $\phi= \delta W / \delta \xi$,
\begin{equation}
\Gamma (\phi,\bar \phi) =  W(\xi,\bar \xi) - \sum_{k,s} \left[
\bar \phi( k,s) \xi (k,s) + \bar \xi( k,s) \phi (k,s) \right]
\label{Gamma}
\, .
\end{equation}
This functional can be expanded in monomials of its sources,
\begin{equation}
\Gamma (\phi,\bar \phi) = \sum_{m \ge 0} \frac{1}{m!}
\sum_{\underline{K}}
\gamma^{(m)} (\underline{K}) \Phi^m(\underline{K}) \, .
\label{Gammaexp}
\end{equation}
$\underline{K}$ is a multi-index which contains $m$
frequencies, wavevectors, spins and Nambu indices of the
$\Phi$-fields. Nambu index $+$ stands for $\bar \phi$ and
$-$ for $\phi$. The $\gamma^{(m)}$ are the $m$-point vertex
functions of the theory. They are antisymmetric with respect
to interchange of two particle coordinates.
 
The two-point vertex $\gamma^{(2)}$ is the inverse of the full
propagator. For spin-rotational and U(1) invariance the
antisymmetric four-point vertex  for incoming particles $k_1,s_1$,
$k_2,s_2$ and outgoing particles $k_3,s_3$, $k_4,s_4$ (the quantum
numbers of particle 4 are dictated by the conservation laws) can
be expressed in terms of a coupling function $V(k_1,k_2,k_3)$ by\cite{salmhofer}
\begin{equation}
\gamma^{(4)}_{s_1s_2s_3s_4} (k_1,k_2,k_3,k_4) =
V(k_1,k_2,k_3) \delta_{s_1s_3}\delta_{s_2s_4} -
V(k_2,k_1,k_3) \delta_{s_1s_4}\delta_{s_2s_3}
\end{equation}

\begin{figure}

\begin{center}

\includegraphics[width=.8\textwidth]{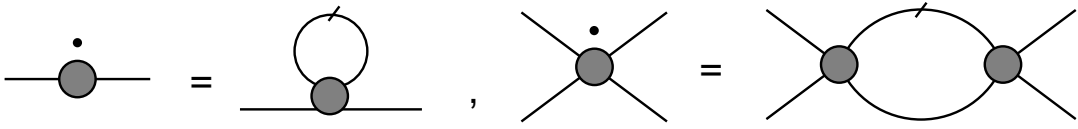}

\end{center}
\caption{RG equations for the two-point and the four-point vertex. The slashed line denotes a single-scale propagator $S_\ell(p)$. The one-loop graph for the four-point vertex $\gamma^{(4)}_\ell$ includes particle-particle and particle-hole contributions. With our truncation $\gamma^{(m)}_\ell=0$ for $m\ge 6$ the feedback of the $\gamma^{(6)}_\ell$ on $\gamma^{(4)}_\ell$ is neglected.}
\label{rgdia}
\end{figure}

Now let us assume that the quadratic part of the action depends
continuously on a parameter $\ell$ 
while the interaction terms does not contain $\ell$. 
Then we can derive an exact equation\cite{wetterich,salmhofer,kopietz} 
for the change of $\Gamma_\ell
(\phi, \bar \phi)$ when $\ell$ is varied by taking the
$\ell$-derivative of Eq. \ref{Gamma}. 
Inserting the field expansion (\ref{Gammaexp}) yields
an infinite hierarchy of differential equations. This set of equations defines
the renormalization group flow of the 1PI vertex function with
respect to the parameter $\ell$. 
In order to make this system tractable, one has
to truncate it at some point. Here all 1PI vertex
functions with $m>4$ are set to zero. 
This is a first approximation that is common
to all RG schemes discussed below. The remaining
equations  for $\gamma^{(2)}_\ell$ and $\gamma^{(4)}_ell$ are shown
graphically in Fig. \ref{rgdia}.

We emphasize that in the exact hierarchy the right hand side contains only one-loop diagrams. This is one of the blessings of the approach which in certain cases even allows for a non-perturbative treatment\cite{berges}.  
Higher-loop diagrams are included via the $\ell$-dependence of the vertices. They can be made explicit by reinserting the integrated flow of the vertices into the right hand side\cite{honedhd}.
The equation for the coupling function
$V_\ell (p_1,p_2,p_3)$ with $p_4=p_1+p_2-p_3$ is\cite{salmhofer}
\begin{eqnarray}
\frac{d}{d\ell} V_\ell (p_1,p_2,p_3) =    {\cal T}_{PP,\ell} + {\cal T}^d_{PH,\ell} +
{\cal T}^{cr}_{PH,\ell} \label{vdot} \end{eqnarray} 
with the one-loop particle-particle contributions ${\cal T}_{PP,\ell}$ and the two different particle-hole channels ${\cal T}^d_{PH,\ell}$ and ${\cal T}^{cr}_{PH,\ell}$ where
\begin{eqnarray}
\lefteqn{  {\cal T}_{PP,\ell} (p_1,p_2;p_3,p_4) = }\nonumber\\
&& - \int dp \, V_\ell ( p_1,p_2,p ) \, L(p,-p+p_1+p_2) \, V_\ell
(p,-p+p_1+p_2 ,p_3) \label{PPdia}
\\[0mm]
 \lefteqn{ {\cal T}^d_{PH,\ell} (p_1,p_2;p_3,p_4) =}  \nonumber \\
&&- \int dp\, \biggl[ -2 V_\ell ( p_1,p,p_3 ) \, L(p,p+p_1-p_3) \,
V_\ell (p+p_1-p_3,p_2,p) \nonumber
 \\ && \qquad +
V_\ell (p_1,p,p+p_1-p_3) \, L(p,p+p_1-p_3) \, V_\ell (p+p_1-p_3,p_2,p)
 \nonumber  \\ && \qquad +
V_\ell ( p_1,p,p_3) \, L(p,p+p_1-p_3) \, V_\ell (p_2,p+p_1-p_3,p)
\biggr]\label{PHddia}
\\[0mm]
\lefteqn{  {\cal T}^{cr}_{PH,\ell}(p_1,p_2;p_3,p_4) =} \nonumber \\
&& - \int dp \, V_\ell (p_1,p+p_2-p_3,p)  \, L(p,p+p_2-p_3) \,
V_\ell (p,p_2,p_3 )\label{PHcrdia}
\end{eqnarray}
In these equations, the product of the two internal lines in the
one-loop diagrams is
\begin{equation} \label{e13}
L(p,p') = S_\ell (p) W^{(2)}_{\ell} (p') + W^{(2)}_{\ell} (p)
S_\ell (p')  \, \end{equation} 
with the so-called single-scale propagator
\begin{equation}
S_\ell (p) = - W^{(2)}_{\ell} (p) \left[ \frac{d}{d\ell} Q_\ell
(p) \right] \, W^{(2)}_{\ell} (p) \, .
\end{equation}
Here, $W^{(2)}_{\ell} (p)$ denotes the full Green's function at RG scale $\ell$. 
 \subsection{Momentum-shell RG, temperature flow and interaction flow}
So far the setup of the fRG equations has been very
general. Now we specify the flow parameter $\ell$. We discuss three
different choices (although many more are possible). The
guiding principle for a good flow parameter is that it should
enable us {\em a)} to approach a specific singularity in the 
perturbation expansion in a controlled way and
{\em b)} to include all other (possibly
also singular) tendencies during the flow. 
Then the RG will give a more realistic picture than perturbation 
expansions which single out one dominant channel. 
In bubble or ladder
summations, the singularities  normally arise due to a pile-up of
logarithms. These are roughly given by $g \log [W/\mathrm{max}(T,\Lambda)]$
with a coupling constant $g$, bandwidth $W$ and lower energy
cutoff $\Lambda$. Hence, if we want to build in such dangerous
terms step by step, we can either vary $\Lambda$, the
temperature $T$, or the coupling $g$. This is made more precise
below.

\paragraph{Cutoff-RG}
Here we introduce a cutoff-function $C_\Lambda [\epsilon(\vec{k})]$
into the quadratic part of the action,
\begin{equation}  Q_\Lambda (k)
= T \sum_{i k_0 ,\vec{k}}  C_\Lambda [\epsilon(\vec{k})] \,
\bar{\psi}_k  \, \left[ -i k_0 + \epsilon (\vec{k})\right]\, \psi_k  \, .
\end{equation}
$C_\Lambda [\epsilon(\vec{k})]$ is very large for $|\epsilon (\vec{k})| \le
\Lambda$ and $C_\Lambda [\epsilon(\vec{k})]=1$ for $|\epsilon (\vec{k})| > \Lambda$  such that modes below $\Lambda$ are not integrated over in the functional integral.
In practice one mainly needs the inverse $C^{-1}_\Lambda
[\epsilon(\vec{k}]$, which can be chosen conveniently as a sharp
step function for analytical manipulations or as a smoothed step
function for numerical treatments. 
The full Green's function is suppressed  for modes with  $|\epsilon (\vec{k})| \le \Lambda$. A frequency
cutoff (or combinations of frequency and band energy) falls into
the same class of cut-off RG. For our $\vec{k}$-space resolved
approach described below a pure energy-cutoff is however most
convenient. The momentum-shell RG  is the
widely used standard\cite{shankar} for many-fermion systems. It can also be derived from other exact RG equations\cite{polchinski,zanchi,tsai,msbook,halboth}.
It approaches Fermi liquid instabilities for fixed model parameters and
fixed temperature. One of the successes of the method in the weakly coupled 2D Hubbard model are the clear signatures for $d$-wave superconductivity over a wide parameter range\cite{zanchi,halboth,hsfr}. 

A serious drawback of the cutoff RG is however the non-uniformity in the RG scale at which one-loop particle-hole (PH) processes with different wavevector transfers $\vec{q}$ are included at low T. 
Typically PH pairs with a particle $\vec{k}$ and a hole $\vec{k}+\vec{q}$ and large $|\vec{q}| \sim k_F $ are integrated out at all scales $\Lambda$ in the flow, depending on how close $\vec{k}$ and $\vec{k}+\vec{q}$  are to the Fermi surface. In contrast with this, PH processes with  small $\vec{q}$
can only occur in the vicinity of the Fermi surface. For $T \to 0$ the support of the PH bubble for $\omega=0,\vec{q} \to 0$ shrinks to a temperature-smeared  $\delta$-function on the FS with width $\sim T$.
Thus, the $\vec{q} \to 0 $ PH-modes are integrated out only for cutoff $\Lambda \le T$, even if the density of states is divergent, and these processes give a singular contribution for $T\to 0$. However, in a coupled flow with a lot of other tendencies, e.g. the omni-presence of Kohn-Luttinger superconducting instabilities, the flow normally diverges {\em before} we get down to $\Lambda \sim T$. The flow has to be stopped and the $\vec{q} \to 0$ PH pairs did not have the chance to contribute to the flow {\em by construction}. Therefore, the cutoff-flow is still biased. Other approaches are needed to study the influence of the $\vec{q} \to 0$ particle-hole excitations. 

The following two schemes work without a cutoff and allow for a uniform inclusion of the various one-loop processes.
\paragraph{Temperature flow}
The temperature-flow scheme\cite{tflow} uses the temperature $T$ appearing  in front of the frequency sum and in the Matsubara frequencies as flow parameter. 
Before the $T$-derivative can be taken, the fermionic
fields need to be rescaled in order to remove the $T^3$ term appearing in front of the interaction term in the
original action. Then the strategy is as follows. The theory is defined at some high T $\sim T_0$ in terms of its
two-point and four-point vertex. 
Here, we simply choose the free propagator and the local
Hubbard repulsion as the initial values at $T_0$. This should be reasonable for sufficiently high $T_0 \sim$ bandwidth, as all
perturbative corrections decay with a negative power of $T$. In the approximate version used below without $\gamma^{(6)}_T$ and self-energy corrections, the right hand side of the flow equation for the interaction is just given by the $T$-derivatives of the one-loop diagrams.

\paragraph{Interaction flow:}
In the interaction flow scheme\cite{ape} we first multiply
$Q$ with a scale factor $1/g$ and split it in two, yielding
\begin{equation}  
Q_g = T \sum_{i \omega_n ,\vec{k}} \bar{\psi}_k g^{-1/2} \, \left[ -i\omega_n + \epsilon (\vec{k})\right] \psi_k g^{-1/2} \, . 
\end{equation}
$g$ will be the flow parameter. We can absorb the factor $1/g$ in
rescaled fields $\tilde{\psi},\bar{\tilde{\psi}}$ defined as $\tilde{\psi}
= g^{-1/2} \psi$. With this the interaction term 
gets an extra factor $g^2$ when written in terms of the new
fields:
\begin{equation}  
V^{(4)}_g = \frac{1}{2N}\sum_{k,k',k+q \atop s,s'} g^2 V(k,k',k+q) \bar{\tilde{\psi}}_{k+q,s}  \bar{\tilde{\psi}}_{k'-q,s'} \tilde{\psi}_{k',s'}  \tilde{\psi}_{k,s} \, . \end{equation}
We observe that changing the scale factor
$1/g$ in $Q_g$ corresponds to changing the strength of the bare
interactions. The rescaled fermions $\tilde{\psi}$, $\bar{\tilde{\psi}}$
describe a system with a bare interaction strength $g^2\,V$. Now 
we can start at $g=0^+$, i.e. at
infinitesimally small bare interaction, and use the flow equations to
integrate up to the desired bare interaction, reached at
$g=1$.  We can also stop the flow at any other value of $g$, with
the functions $g \Sigma$ and  $g^2 V_g(k,k',k+q)$ being the self-energy 
and interacting vertex function for the bare interaction
$g^2 V (k,k',k+q)$.
We call this the {\em interaction flow} (IF). Related schemes have been proposed by Polonyi\cite{polonyi} and Meden\cite{meden}.

In the IF, singularities on the right hand side of the flow equation are not
regularized by the flow parameter.  Thus, the IF
scheme has to be performed at $T>0$, when the
individual one-loop diagrams are bounded. The strength of the bare 
interaction is increased continuously, and in the course of the
interaction flow potential singularities are approached from
below.\\[1mm]

\begin{figure}

\begin{center}

\includegraphics[width=.8\textwidth]{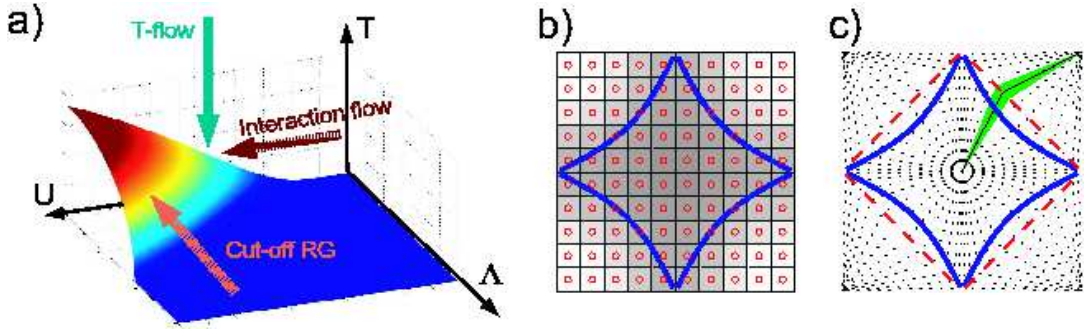}

\end{center}
\caption{ a) The different fRG approaches in the
parameter space spanned by interaction strength $U$, temperature
$T$ and infrared cutoff $\Lambda$. The surface represents the
critical manifold below which perturbation theory diverges.
 b) Box discretization of the BZ and the Fermi surface. c) $N$-patch discretization. The patches extend from the origin across the FS to the corners of the BZ. }
\label{RGschemes}

\end{figure}

Fig.\ \ref{RGschemes} a) illustrates how the various methods detect perturbative singularities from different
directions in parameter space.
We emphasize that the $T$-flow and the IF scheme do not correspond to
viewing a system on different length scales.

\section{Application to the 2D Hubbard model}
Here we review results\cite{honedhd,hsfr} obtained for the 2D $t$-$t'$ Hubbard model
with next-nearest neighbor hopping $t' \not=0$. 
The case $t'=0$ was studied with related cutoff-RG schemes e.g. by Zanchi and Schulz\cite{zanchi} and Halboth and Metzner\cite{halboth}.

\subsection{Calculational issues and approximations} 
All fRG approaches to the Hubbard model have so far neglected the 
frequency dependence of the interaction vertex by calculating the static part and approximating all other vertices with the static one. 
This can be argued to be good as the initial interactions are frequency independent, and the corrections to the leading divergences in the static components are small by power counting. 
The next approximation is the neglect of the self-energy corrections. The consistency of this can be checked by calculating the self-energy from the flow of the couplings. 
With these approximations and the initial truncation of the hierarchy one is left with the flow of the coupling function 
$V_\ell(\vec{k}_1,\vec{k}_2,\vec{k}_3)$. 
This function of three wavevectors can either be discretized directly  on a grid in the Brillouin zone (see Fig. \ref{RGschemes} b)), or, more physically motivated, in the so-called $N$-patch scheme, first used in this context by Zanchi and Schulz\cite{zanchi}. This scheme takes advantage of the fact that for the standard Fermi liquid instabilities the wavevector dependence of the coupling function $V(\vec{k}_1,\vec{k}_2,\vec{k}_3)$ has a relevant part, which is the dependence of the processes between different points $\vec{k}^F_1,\vec{k}^F_2,\vec{k}^F_3$ on the FS when the $\vec{k}^F_i$ are varied around it. Furthermore there is an irrelevant part, which is the dependence in the directions orthogonal to the FS. Hence, one calculates the $V(\vec{k}_1,\vec{k}_2,\vec{k}_3)$  for $\vec{k}_1,\vec{k}_2,\vec{k}_3$ on the FS and treats it as piecewise
 constant when $\vec{k}_1,\vec{k}_2$, and $\vec{k}_3$ move within elongated patches stretching from the origin of the BZ to the $(\pm \pi,\pm \pi)$-points 
(see Fig. \ref{RGschemes} c)).

Then, the RG flow is started at an initial scale $\Lambda_0$ for the cutoff RG, at temperature $T_0$ for the $T$-flow, or initial interaction strength $g^2= 0^+$ for the interaction flow, respectively. The initial interaction is chosen as $V_\ell (\vec{k}_1,\vec{k}_2,\vec{k}_3)=U$ ($=g^2U$ for the IF).
What is typically encountered at low $T$ is a {\em flow to strong coupling}, where for a certain flow parameter $\ell_c$ one or several components of $V_\ell (\vec{k}_1,\vec{k}_2,\vec{k}_3)$ become large. At that point the approximations break down, and the flow has to be stopped. 
Physical information is obtained from susceptibilities that can be calculated from the flow and directly from the $\vec{k}$-dependence of $V_\ell (\vec{k}_1,\vec{k}_2,\vec{k}_3)$. For example, if only processes with $\vec{k}_1+\vec{k}_2=0$ become large in magnitude (typically strongly attractive), the flow to strong coupling corresponds to a Cooper instability, indicating a dominant tendency to form a superconducting state in the parameter regime beyond the divergence. The symmetry of the superconducting pairing can be read off from the $\vec{k}_1$-dependence of the coupling function.

\subsection{Flow at van Hove filling}
First we consider the flow at the so-called van Hove (VH) filling, where the FS is kept fixed at the VH singularities of the free dispersion at $\vec{k}=(\pi,0)$ and $(0,\pi)$. We change the band filling by varying the next-nearest neighbor hopping $t'$. 

Using the $T$-flow scheme one finds three regimes\cite{tflow}. In the first regime for small $|t'|$ the FS is properly nested and there is a antiferromagnetic (AF) spin density wave regime. Then, for increasing $|t'|$, the $d_{x^2-y^2}$-wave superconducting tendencies get stronger and there is a regime where the corresponding susceptibility diverges most strongly. Around $t' = - 0.33t$, the flow to strong coupling is suppressed by  orders of magnitude. 
On the other side of this potential quantum critical point the flow is dominated by ferromagnetic ordering tendencies. The same qualitative picture is found with the IF scheme\cite{ape}, although in this approach the $d$-wave tendencies seem to be weaker by a certain degree, most probably due to the unfavorable box-discretization (Fig. \ref{RGschemes} b)) used. 
Also, a  parquet approach focusing on the VH regions gives the same order of instabilities\cite{katanin-pq}.
On the contrary, the cutoff RG misses the ferromagnetic regime completely. The reason is the above-mentioned inability of the cutoff RG to treat small-$\vec{q}$ particle-hole pairs on equal footing with other tendencies. Instead it produces a low-$T_c$ Kohn-Luttinger instability in the singlet channel. 
 
\begin{figure}
\begin{center}
\includegraphics[width=.7\textwidth]{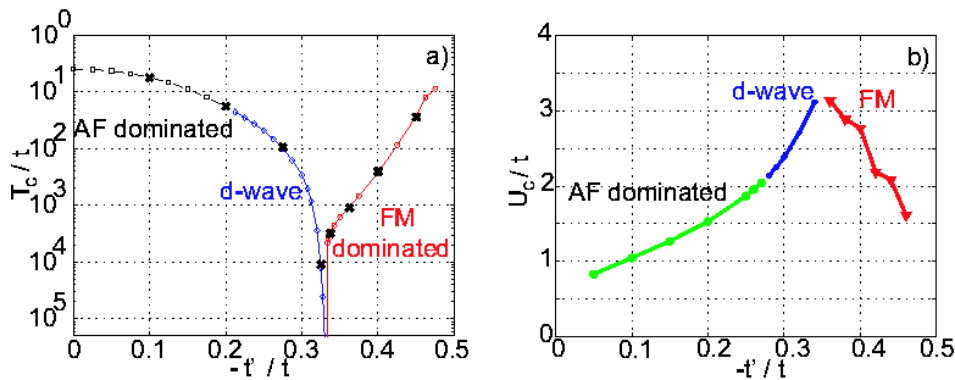}
\end{center}
\caption{a) Critical temperatures $T_c$ for the flow to strong coupling and dominant correlations at the van Hove filling, obtained with $T$-flow RG for $N=48$ patches and $U=3t$.
b) Critical bare interaction strengths for the flow to strong coupling, obtained  with IF scheme at $T=0.001t$ and the van Hove filling and $12\times 12$ box discretization.
}
\label{vhpd}

\end{figure}

\subsection{Flow with fixed $t'$: saddle point regime}
Next we briefly sketch the tentative phase diagrams from the RG flow for fixed $t'=-0.3t$ as obtained with the cutoff RG\cite{hsfr,honedhd}. These values of $t'$ might be relevant for the high-$T_c$ problem. However, the Mott insulating state of the high-$T_c$ compounds can not be obtained with our weak coupling approach. Hence, our results cannot be directly compared to the cuprates. We can however read off some tendencies. 
For these parameter values there are still quantitative differences between the various RG schemes. These mainly concern the location of the boundaries and the numbers for critical scales. The order of the regimes however is the same for all approaches.

\begin{figure}

\begin{center}
\includegraphics[width=.7\textwidth]{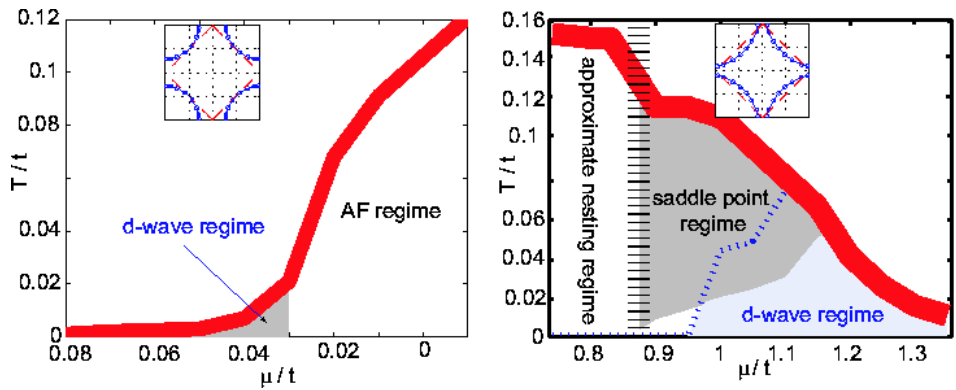}

\end{center}

\caption{Summary of the RG flow for fixed $t'=-0.3t$ 
obtained with $N$-patch cutoff RG. 
The left plot is for electron doping, i.e. band filling $n>1$, while the right plot is for hole doping, $n<1$. 
The band is half filled for $\mu \approx -0.65t$.  
The insets show the location of the Fermi surfaces for the different chemical potentials. Above the thick lines the flow to strong coupling is cut off by temperature. 
}
\label{pdtp3}

\end{figure}

First let us consider band fillings $n>1$ which correspond to electron-doping in the high-$T_c$ language. Here the RG finds just two phases when $n$ is varied. One is an AF nesting regime with relatively large critical scales closer to half filling. Increasing the particle number reduces nesting, and finally $(\pi,\pi)$ particle-hole become impossible at low $T$ (see FS in the left plot of Fig. \ref{pdtp3}). Then (or at somewhat lower doping, depending on the RG scheme) the AF tendencies become cut off. What remains is a low-scale $d$-wave superconducting instability. 

On the hole-doped side with $n<1$ the FS is close to the VH points and the {\em saddle point regions} around $({\pi,0})$ and $({0,\pi})$ play a dominant role. Between an AF nesting regime near half filling and a $d$-wave regime further away we find another regime, the so-called {\em saddle point regime}. Here several channels, most prominently the AF spin density wave, the $d$-wave Cooper, and also the $d$-density wave channel grow with comparable strength. All these channels reinforce each other mutually\cite{hsfr}. 
Hence, it is not obvious that the system goes into a long range ordered phase. 
A similar flow arises in Hubbard two-leg ladder systems. There the ground state  is known to be a short-range correlated Mott insulator. 
Therefore it is conceivable that the saddle point regions of the 2D system will adopt a similar gapped strong coupling state with a truncated FS\cite{furukawa,hsfr}. We have elaborated on this idea recently by exact diagonalization of effective Hamiltonians suggested by the RG flow\cite{laeuchli}. 

\section{Continuation of the flow through the instability}
So far we have only discussed RG flows where the effect of self-energy corrections has been neglected. This approximation will certainly break down when the couplings get large and it is desirable to include at least parts of the self-energy.

There have been several fRG studies\cite{honedhd,zanchiZ,zflow,katanin-sigma,rohe} which gave some insights how the normal state self-energy is altered by the flow of the interactions. Yet in these works the RG flow still had to be stopped when the coupling constants became large. 
Physically one may expect that in many cases the opening of a gap in the fermionic spectrum should regularize this apparent divergence. 
Recently we have analyzed this process in the reduced BCS model using the 1PI cutoff-RG\cite{gapflow}. In this model only particle pairs with zero total momentum interact with an attractive coupling constant. A theory with two- and four-point vertices only can be seen to be exact. 
The essential step is to include a small symmetry-breaking off-diagonal part $\Delta_0$ into the initial condition for the self-energy. Now also anomalous propagators appear in the flow when the cutoff is lowered. They generate a anomalous interaction vertex which has 4 incoming or 4 outgoing lines.
The flow equations for the self-energy and the interaction vertices as shown in Fig. \ref{rgdia} can be integrated down to cutoff $\Lambda =0$ without a divergence of the interactions. A necessary point is a reorganization of the flow for the four-point vertex proposed recently by Katanin\cite{kat}. The $\Lambda=0$-result for the anomalous self-energy approaches the BCS result for $\Delta_0 \to 0$. With initial $\Delta_0>0$ the flow of the four-point vertex remains finite. Only for $\Delta_0 \to 0$, a combination of normal and anomalous vertices diverges like $1/\Delta_0$. This corresponds to the vanishing mass of the Goldstone boson when the U(1)-symmetry is restored.

So far this scheme has only been applied to the reduced BCS model. It will however also work for more general cases where the interplay of particle-particle and particle-hole processes drives the system towards a superconducting instability. If a nonzero $\Delta_0$ is included in the initial conditions, the flow down to $\Lambda =0$ will converge to a state with a gapped spectrum and a controlled magnitude of the interactions. The strength of this scheme is that no mean-field
decoupling of the interactions is required. In particular, the attractive interaction which goes into the equation for the anomalous self-energy is generated automatically during the flow to all orders in the bare interaction.
This should allow for a much more precise investigation of the phase diagrams of weakly interacting fermion systems.

\section*{Acknowledgments} I thank Manfred Salmhofer for
sharing many ideas with me and T. Maurice Rice for motivation. 
Daniel Rohe is acknowledged for comments on the manuscript.

\end{document}